\documentclass[prd,preprintnumbers,amsmath,amssymb]{revtex4}
\usepackage{epsfig}
\usepackage[latin1]{inputenc}
\begin{document} 
\title{Testing parameters in an eikonalized dynamical gluon mass model\footnote{Presented by D. A. Fagundes at \textit{LISHEP 2011 - Workshop on LHC - Present and Future}, Rio de Janeiro, RJ, July 05-10, 2011, Brazil.}}
\author{D.A. Fagundes$^{1}$, E.G.S. Luna$^{2}$, M.J. Menon$^{1}$, A.A. Natale$^{3}$\\\vspace*{0.8cm} } 
\affiliation{$^{1}$Instituto de Física Gleb Wataghin,
Universidade Estadual de Campinas, UNICAMP 
13083-859 Campinas SP, Brazil\\
$^{2}$Instituto de Física, Universidade Federal do Rio Grande do Sul, Caixa Postal 15051, CEP 91501-970, Porto Alegre, RS, Brazil. \\
$^{3}$Instituto de Física Teórica, UNESP - Universidade Estadual Paulista, Rua Dr. Bento T. Ferraz, 271, Bloco II
01140-070, São Paulo - SP, Brazil}

\begin{abstract}
In the framework of a dynamical gluon mass model recently developed,
we investigate the effects of two essential parameters in the description
of elastic $pp$ and $\bar{p}p$ data at high energies: the soft Pomeron
intercept and the dynamical gluon mass. By considering relevant
numerical intervals for both parameters and fits to the experimental data
up to 1.8 TeV, with good statistical results, we discuss the predictions of
the physical quantities at the LHC energies (7 and 14 TeV). We conclude that
these quantities are sensitive to those variations
and the predictions are correlated with the intervals considered
for both parameters. This conclusion puts limits on the reliability
of QCD inspired models predictions at the LHC energies, mainly those models
with ad hoc fixed  values for the mass scale and
the Pomeron intercept. 
\end{abstract}

\maketitle


\section{Introduction}
The TOTEM experiment has been designed to study elastic and diffractive scattering at LHC, providing crucial information on the
$pp$ total cross section  and the elastic differential cross section at energies 7-14 TeV. From the theoretical point of view, elastic scattering still constitutes an open problem for QCD since perturbative techniques can only be applied 
in limited kinematical regions and a nonperturbative approach for the soft scattering processes is still lacking. Moreover, 
a wide variety of phenomenological models present good descriptions of the existing data, however with different physical pictures. In
this respect the expected TOTEM data will be very important in the selection of models. In particular, the class of models referred to
as ``QCD inspired'' plays an important role in this scenario, since they are based on some inputs directly connected with QCD.\\
In this work we analyze elastic scattering in the context of a QCD inspired model, with a dynamical gluon mass
used as regulator for the infrared region, which we denote DGM (Dynamical Gluon Mass) Model \cite{luna01}. 
We discuss here two novel aspects: (i) new developments in the original formulation; 
(ii) the influence of a mass scale and the intercept of soft Pomeron in the predictions at the LHC energy region.\\
This note is organized as follows:  In section 2 we review the general structure of the DGM model and its main inputs. 
In section 3 we present the fit results on $pp$ and $\bar{p}p$ total cross sections, $\rho$ parameter and elastic differential cross sections. 
Our conclusions are the content of section 4.

\section{DGM Model}
The DGM model is largely based on the eikonal approach previously discussed by M. Block and collaborators \cite{block01,block02},
but with two novel physically motivated ingredients for the infrared mass scale and the running coupling constant, both connected
to an infrared dynamical gluon mass scale. In this section we shortly review the main formulas and the new developments in the original
formulation.

\subsection{Dynamical Gluon Mass}
Cornwall\cite{cornwall} has shown that in obtaining a gauge invariant solution to
the Schwinger-Dyson equation for the gluon propagator, a dynamical gluon mass arises naturally. In doing so he was able to obtain a 
nonperturbative expression for the running coupling constant, which depends on a dynamical gluon mass and freezes at the IR 
region. In DGM model we will account for nonperturbative effects in elastic scattering considering Cornwall's
expressions for the coupling constant
\begin{eqnarray}
\bar{\alpha}_{s} (\hat{s})= \frac{4\pi}{\beta_0 \ln\left[
(\hat{s}+4M_g^2(\hat{s}))/\Lambda^2 \right]},
\label{acor}
\end{eqnarray}
and for the dynamical gluon mass,
\begin{eqnarray}
M^2_g(\hat{s}) = m_g^2 \left[ \frac{\ln \left( \frac{\hat{s}+4{m_g}^2}{\Lambda^2} \right) }{\ln
\left( \frac{4m_g^2}{\Lambda^2} \right) } \right]^{- 12/11} ,
\label{mdyna}
\end{eqnarray}
where $\beta_{0} = 11 - \frac{2}{3}n_{f}$ ($n_{f}$ is the number of flavors), $\Lambda = \Lambda_{QCD}$ and $\tau = \hat{s}/s$ represents the fraction of 
energy carried by partons inside the colliding hadrons and m$_{g}$ is the gluon mass scale. In this work we have considered $n_{f} = 4$ and $\Lambda = 284$ MeV.

\subsection{Eikonal and Impact Parameter Representations}
In the eikonal representation the elastic scattering amplitude, $A(s,t)$, is written as (azimuthal symmetry assumed):
\begin{equation}
 A(s,t) = i \int{bdb}J_{0}(qb)[1-\mathrm{e}^{i\chi(s,b)}],
\end{equation}
where $q^{2} = -t$ and $\chi(s,b) = \chi_{R}(s,b)+i \chi_{I}(s,b)$ is the complex eikonal function, expressed  
in terms of even/odd contributions for $pp$ and $\bar{p}p$ scattering:
\begin{equation}
\chi_{pp}^{\bar{p}p}(s,b) = \chi^{+} (s,b) \pm \chi^{-} (s,b). 
\end{equation}
The total cross section, $\sigma_{tot}$, $\rho$ parameter and the differential elastic cross section, $d\sigma_{el}/dt$ , are given in terms of $\chi(s,b)$
by
\begin{eqnarray}
\sigma_{tot}(s) &=& 4\pi \int_{_{0}}^{^{\infty}} \!\! b\, db\, [1-e^{-\chi_{_{I}}(b,s)}\cos \chi_{_{R}}(b,s)],
\label{degt1} \\\vspace{0.5cm}
\rho (s) &=& \frac{\textnormal{Re} \{ i \int b\, db\, [1-e^{i\chi (b,s)}]  \}}{\textnormal{Im} \{ i \int b\,
db\, [1-e^{i\chi (b,s)}]  \}},
\label{degthyj1}\\\vspace{0.5cm}
\frac{d\sigma_{el}}{dt} (s,t) &=& \pi| \int{bdb}J_{0}(qb)[1-e^{i\chi(b,s)}]|^2.
\end{eqnarray}

\subsection{Elementary Processes}

In the DGM model we use as input for the odd eikonal \cite{luna01,block02}
\begin{eqnarray}
\chi^{-} (b,s) = C^{-}\, \Sigma \, \frac{m_{g}}{\sqrt{s}} \, e^{i\pi /4}\, 
W(b;\mu^{-}),
\label{bl2}
\end{eqnarray}
where $\Sigma \equiv \frac{9\pi \bar{\alpha}^2_{s}(0)}{m_{g}^2}$. This ``instrumental'' expression accounts
for differences between $pp$ and $\bar{p}p$ channels at low energies. The even part contribution is connected with elementary interactions, 
gluon-gluon, quark-gluon, quark-quark, as follows:
\begin{eqnarray}
\chi^{+}(b,s) &=& \chi_{qq} (b,s) +\chi_{qg} (b,s) + \chi_{gg} (b,s) \nonumber \\
&=& i [\sigma_{qq}(s) W(b;\mu_{qq}) + \sigma_{qg}(s) W(b;\mu_{qg})\nonumber\\
&+& \sigma_{gg}(s) W(b;\mu_{gg})] ,
\end{eqnarray}
where $W(b,\mu_{ij})$, $ij = qq, qg, gg$, is the overlap function in the impact parameter space and $\sigma_{ij}$ represents elementary
cross sections of interactions between quarks and gluons.\\
We parametrize the $qq$ and $qg$ contributions based on the energy dependence originated from approximate forms of the distribution functions
of quarks and gluon at small $x$ region\cite{block02}. Therefore, the eikonals $\chi_{qq}$ and $\chi_{qg}$ are given by
\begin{eqnarray}
\chi_{qq}(b,s) &=& i \, \Sigma \, A \,
\frac{m_{g}}{\sqrt{s}} \, W(b;\mu_{qq}),\\
\label{mdg1} 
\chi_{qg}(b,s) &=& i \, \Sigma \left[ A^{\prime} + B^{\prime} \ln \left( \frac{s}{m_{g}^{2}} \right) \right] \,
W(b;\mu_{qg}).
\label{mdg2}
\end{eqnarray}
The $gg$ contribution is written as
\begin{eqnarray}
\sigma_{gg}(s) = C^{\prime} \int_{4m_{g}^{2}/s}^{1} d\tau \,F_{gg}(\tau)\,
\hat{\sigma}_{gg} (\hat{s}) ,
\label{sloh1}
\end{eqnarray}
where $F_{gg}(\tau)$ is the gluon distribution function,
\begin{eqnarray}
F_{gg}(\tau)=[g\otimes g](\tau)=\int_{\tau}^{1} \frac{dx}{x}\, g(x)\,
g\left( \frac{\tau}{x}\right),\label{fgg}
\end{eqnarray}
and $\hat{\sigma}_{gg}(\hat{s})$ represents the $gg\rightarrow gg$ nonperturbative cross section  
\begin{eqnarray}
\hat{\sigma}_{gg}(\hat{s}) &=& \left(\frac{3\pi \bar{\alpha}_{s}^{2}}{\hat{s}}\right) \left[\frac{12\hat{s}^{4}
- 55 M_{g}^{2} \hat{s}^{3} + 12 M_{g}^{4} \hat{s}^{2} + 66 M_{g}^{6} \hat{s} -
8 M_{g}^{8}}{4 M_{g}^{2} \hat{s} [\hat{s} - M_{g}^{2}]^{2}}\right] \nonumber\\
&-& \left(\frac{3\pi \bar{\alpha}_{s}^{2}}{\hat{s}}\right) \left[3 \ln \left( \frac{\hat{s} - 3M_{g}^{2}}{M_{g}^{2}}\right)\right].
\label{h16}
\end{eqnarray}
analogous to the one found in \cite{luna01}, but instead of $m_{g}$ we have considered explicitly the energy 
dependence in $M_{g}(\hat{s})$. 
Moreover, in eq. (\ref{fgg}) we have set the following gluon distribution function (for the small $x$ region): 
\begin{eqnarray}
g(x) = N_{g} \, \frac{(1-x)^5}{x^{J}},
\label{distgf}
\end{eqnarray}
were $N_{g} = \frac{1}{240}(6-\epsilon)(5-\epsilon)...(1-\epsilon) $, $J=1+\epsilon$ and $\epsilon$ is the soft Pomeron 
intercept.\\
The nonperturbative cross section, $\sigma_{gg}(s)$, drives the total cross
section, $\sigma_{tot}(s)$, in high energies (typically starting from 1 TeV). Therefore, 
the asymptotic behavior of $\sigma_{tot}(s)$ is related to the one of $\sigma_{gg}(s)$
\begin{eqnarray}
\lim_{s\rightarrow \infty} \int_{4m_{g}^{2}/s}^{1} d\tau \,F_{gg}(\tau)\,
\hat{\sigma}_{gg} (\hat{s}) \sim \left(\frac{s}{4m_{g}^{2}}\right)^{\epsilon}\ln \left(\frac{s}{4m_{g}^{2}}\right)
\end{eqnarray}

This result indicates the dependence of $\sigma_{tot}$ on m$_{g}$ and $\epsilon$ at high energies and that is the point we
are interested to quantitatively investigate here.

\section{Fits and Results}
As first steps in this study we analyze narrow intervals of m$_{g}$ and $\epsilon$ considering 3 values in each case:
\begin{center}
m$_{g}$: 350, 400 and 450 MeV\\
$\epsilon$: 0.080, 0.085 and 0.090
\end{center}
The range of masses are compatible with recent studies on the structure functions F$_{2}$ at small $x$ \cite{luna02} and that
of $\epsilon$ with the analysis of bounds for the soft Pomeron intercept \cite{lmm}.  
\subsection{Fit Procedure}
For each pair m$_{g}$, $\epsilon$ and through eqs. (5-7) we have developed simultaneous fits to the available data on $pp$ and $\bar{p}p$ forward observables
$\sigma_{tot}$ and $\rho$ above 10 GeV and to $\bar{p}p$ differential elastic cross section at 546 GeV 
and 1.8 TeV.  To minimize the data we used the class TMinuit of the CERN ROOT Framework \cite{root} and the MIGRAD minimizer setting a 
confidence level of 90\%. In the first step we fixed the gluon mass and changed the value of the soft Pomeron intercept in the above range. 
Secondly, we fixed the intercept and variate the gluon mass.

\subsection{Results}
We present here some representative results in the following cases: 

\begin{center}
\hspace{0.5cm}m$_{g}$=400 MeV and $\epsilon$: 0.080, 0.085 and 0.090\\
$\epsilon$=0.080 and m$_{g}$: 350, 400 and 450 MeV.
\end{center}
The statistical information on the fit results are shown in Table 1 and the values of the free parameters in the case $\epsilon$=0.080
in Table 2. Fit results and predictions are displayed in Figs. 1 (m$_{g}$=400 MeV) and 2 ($\epsilon$=0.080) and in Tables 3 and 4.  

\begin{center}
 \small{Table 1. Statistical information on the fit results.}\\\vspace{0.3cm}
\begin{tabular}{|c|cccc|}
\hline
m$_{g}$ = 400 MeV& $\epsilon$:  & 0.080 & 0.085 & 0.090\\
&$\chi^{2}$/DOF: & 0.95 & 0.96 & 0.96\\
\hline 
$\epsilon$ = 0.080& m$_{g}$ (MeV): & 350 & 400 & 450 \\ 
&$\chi^{2}$/DOF & 0.96 & 0.95 & 0.96\\
\hline
\end{tabular} 
\end{center}

\vspace{1cm}

\begin{center}
\small{Table 2. Values of the fit parameters for $\epsilon$ = 0.080. $C', A, A', B, C^{-}$ are dimensionless 
and $\mu_{ij}, ij = qq, qg$ and $gg$ have dimension of GeV.}\\\vspace{0.3cm}
\begin{tabular}{c|c|c|c}
\hline\hline
m$_{g}$ (MeV):           &350   & 400 & 450 \\ \hline
$C'$         &($2.336 \pm 0.039$)x10$^{-3}$ &($3.79 \pm 0.17$)x10$^{-3}$& ($5.00 \pm 0.11$)x10$^{-3}$ \\
$\mu_{gg}$   &$0.6611 \pm 0.0048$&$0.651 \pm 0.066$& $0.6353 \pm 0.0074$\\
$A$          &$7.02 \pm 0.22$&$10.7 \pm 1.4$& $21.30 \pm 0.71$\\
$\mu_{qq}$   &$1.478 \pm 0.023$&$1.9841 \pm 0.0038$& $1.537 \pm 0.064$\\
$A'$         &$0.5473\pm0.0018$&$0.874\pm0.059$& $1.0128\pm0.0066$  \\
$B'$         &($2.017 \pm 0.010$)x10$^{-2}$&($4.51 \pm 0.62$)x10$^{-2}$& ($11.30 \pm 0.10$)x10$^{-2}$  \\
$\mu_{qg}$   &$0.8414 \pm 0.0027$&$0.8361 \pm 0.0019$& $0.8261 \pm 0.0027$\\
$C^{-}$      &$1.898 \pm 0.039$&$3.03 \pm 0.40$& $3.981 \pm 0.091$  \\
$\mu^{-}$    &$0.2841 \pm 0.0025$&$0.41 \pm 0.17$& $0.263 \pm 0.065$\\
\hline\hline  
\end{tabular}
\end{center}
\vspace{1cm}

\begin{center}
\vspace{0.5cm}
\includegraphics*[width=8cm,height=7cm]{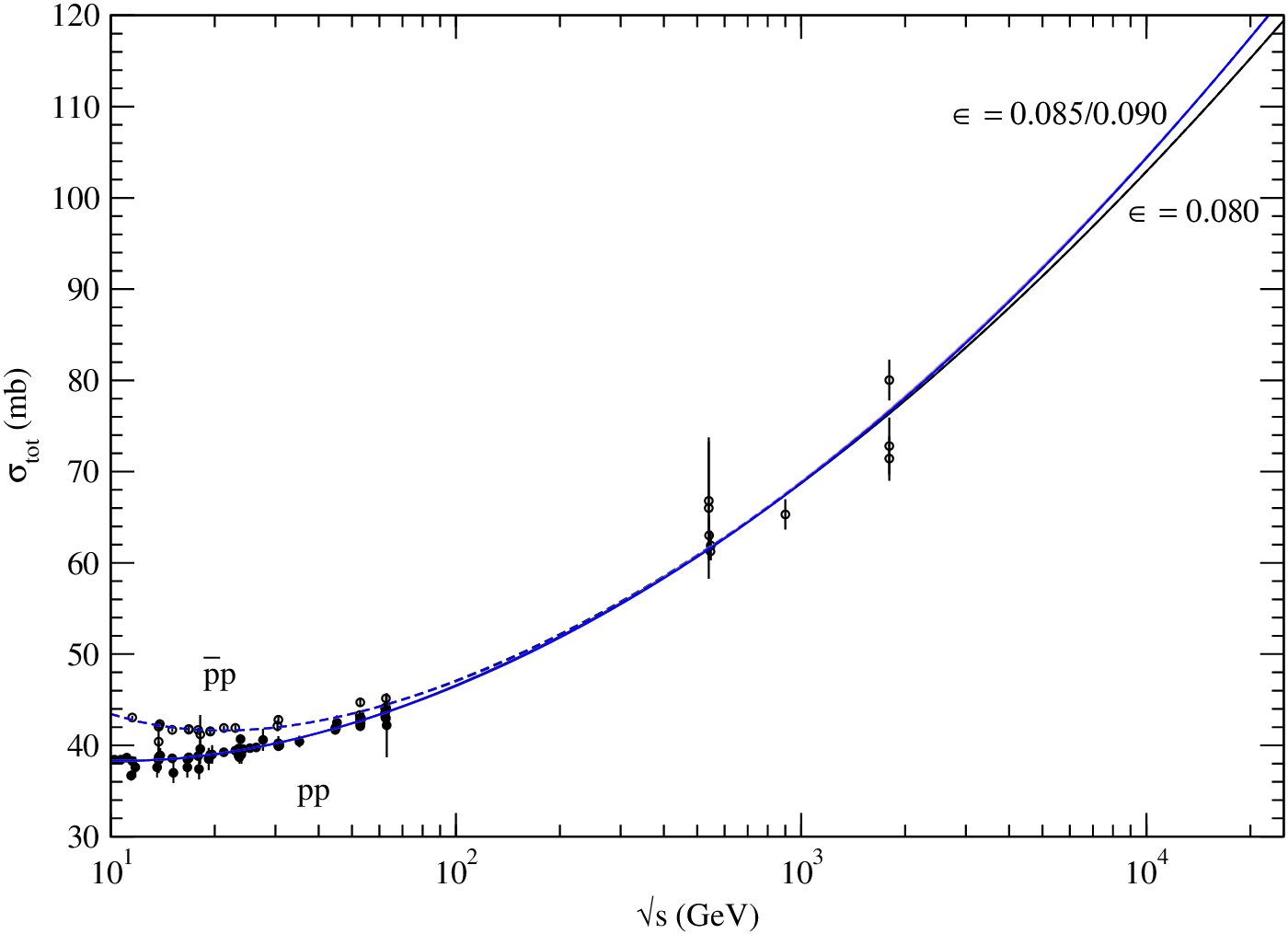}\hspace*{0.2cm}
\includegraphics*[width=8cm,height=7cm]{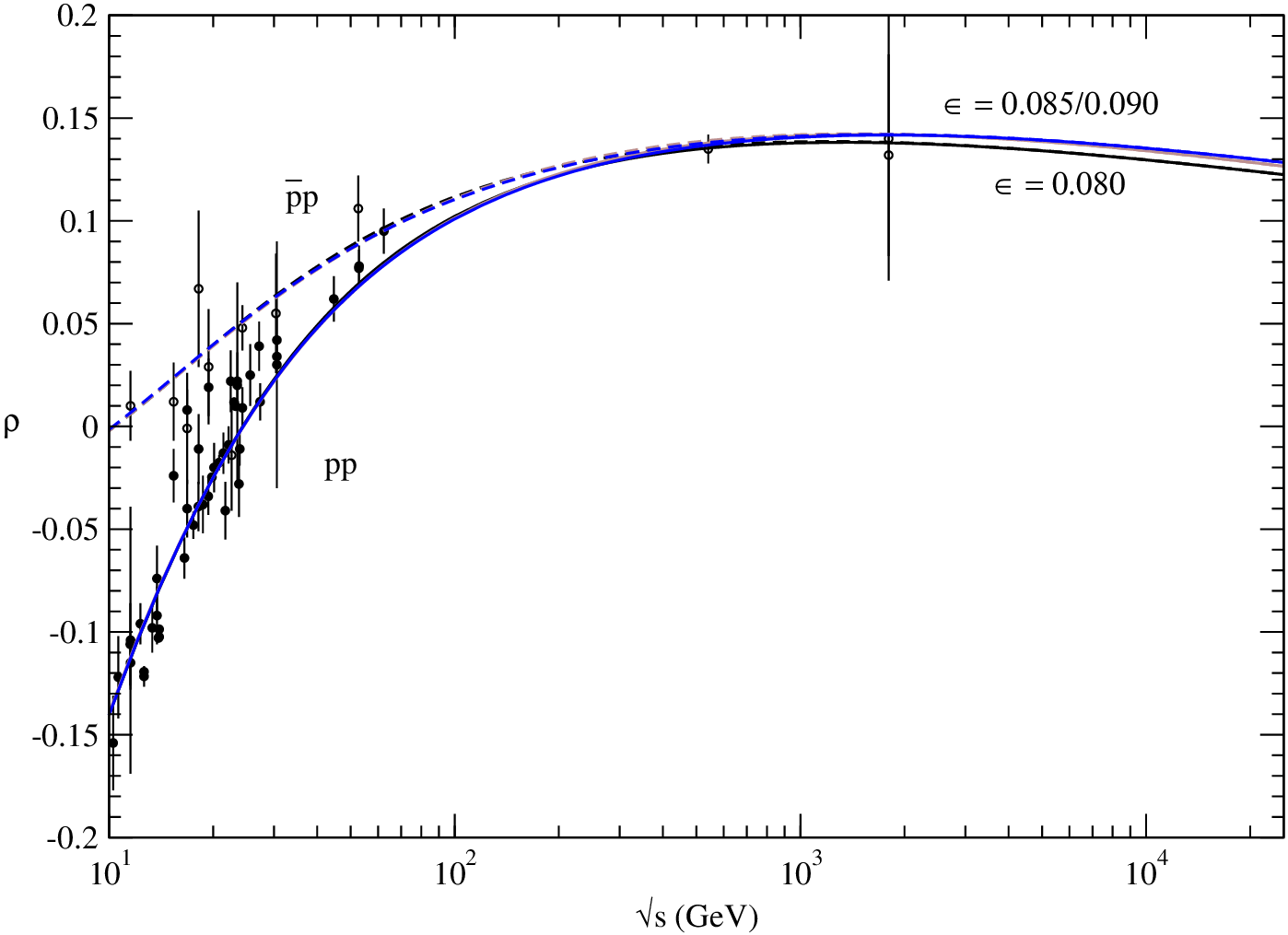}\vspace{0.5cm}
\includegraphics*[width=8cm,height=7cm]{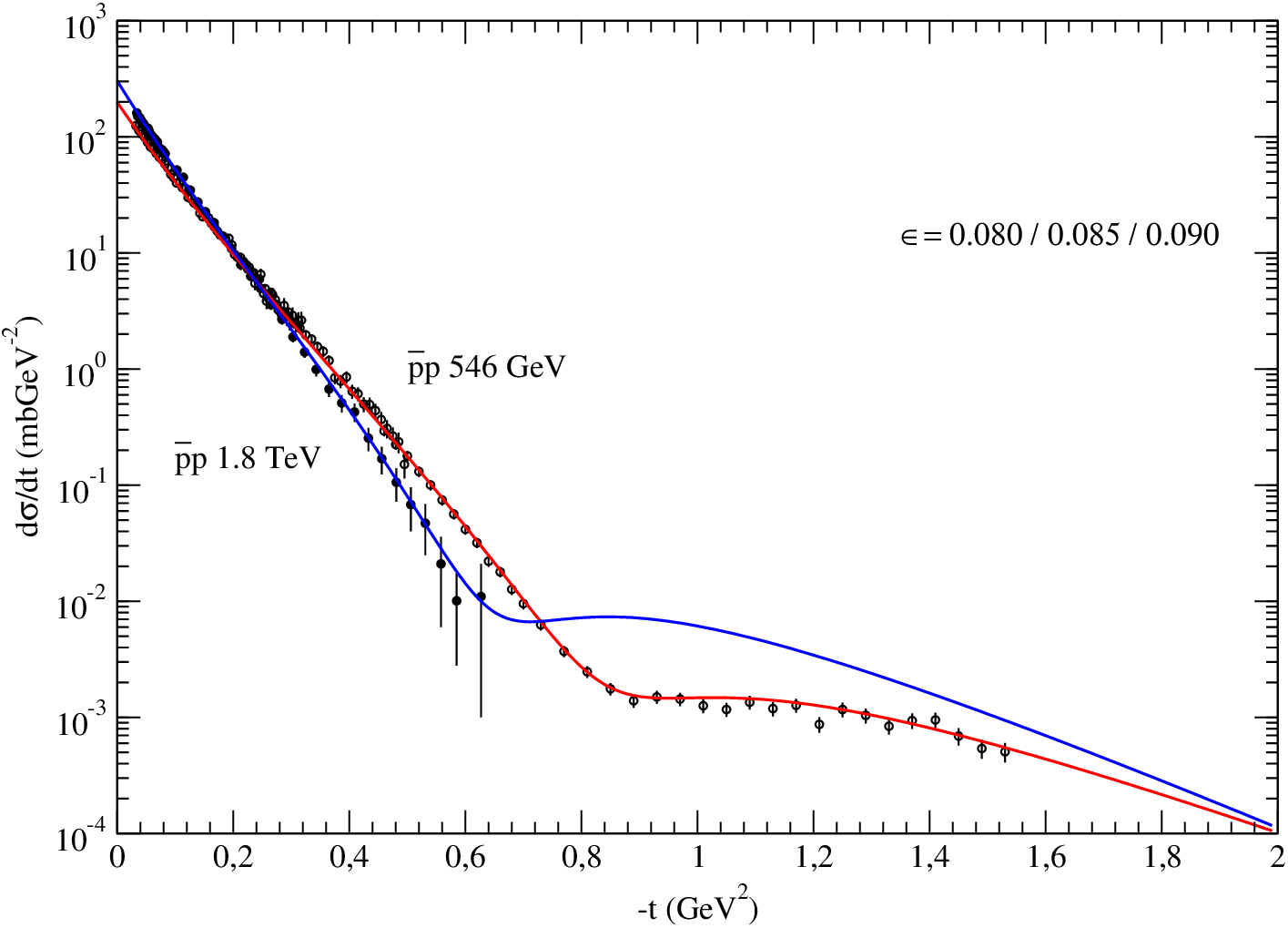}\hspace*{0.2cm}
\includegraphics*[width=8cm,height=7cm]{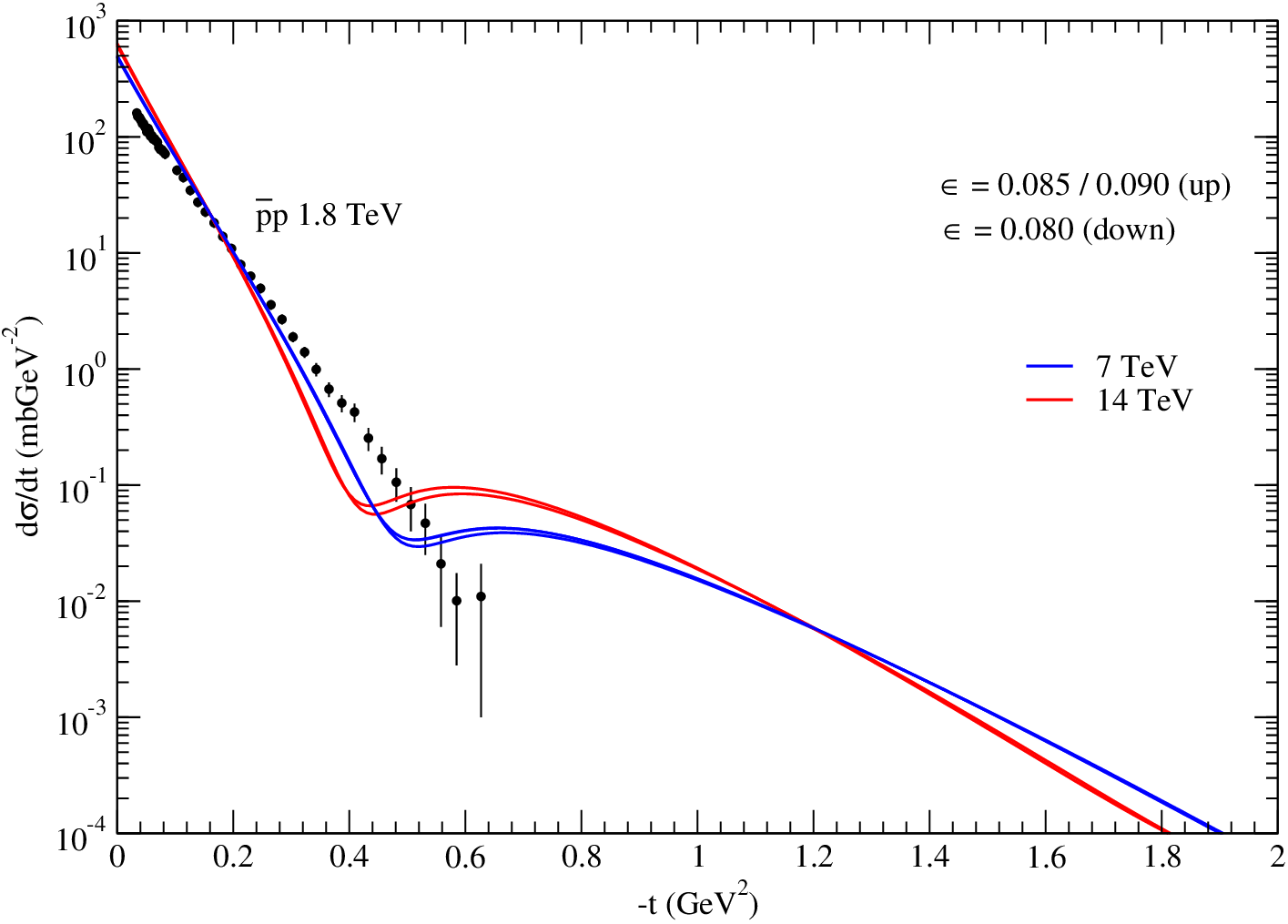}
\end{center}
{\small Fig. 1: Simultaneous fits to $\sigma_{tot}$, $\rho$ and $d\sigma_{el}/dt$ (546 GeV and 1.8 TeV) with $m_{g}$ = 400 MeV and $\epsilon$ in the range
0.08 $-$ 0.09; also shown the predictions for LHC at 7 and 14 TeV.}\\\\

\begin{center}
\vspace{0.5cm}
\includegraphics*[width=8cm,height=7cm]{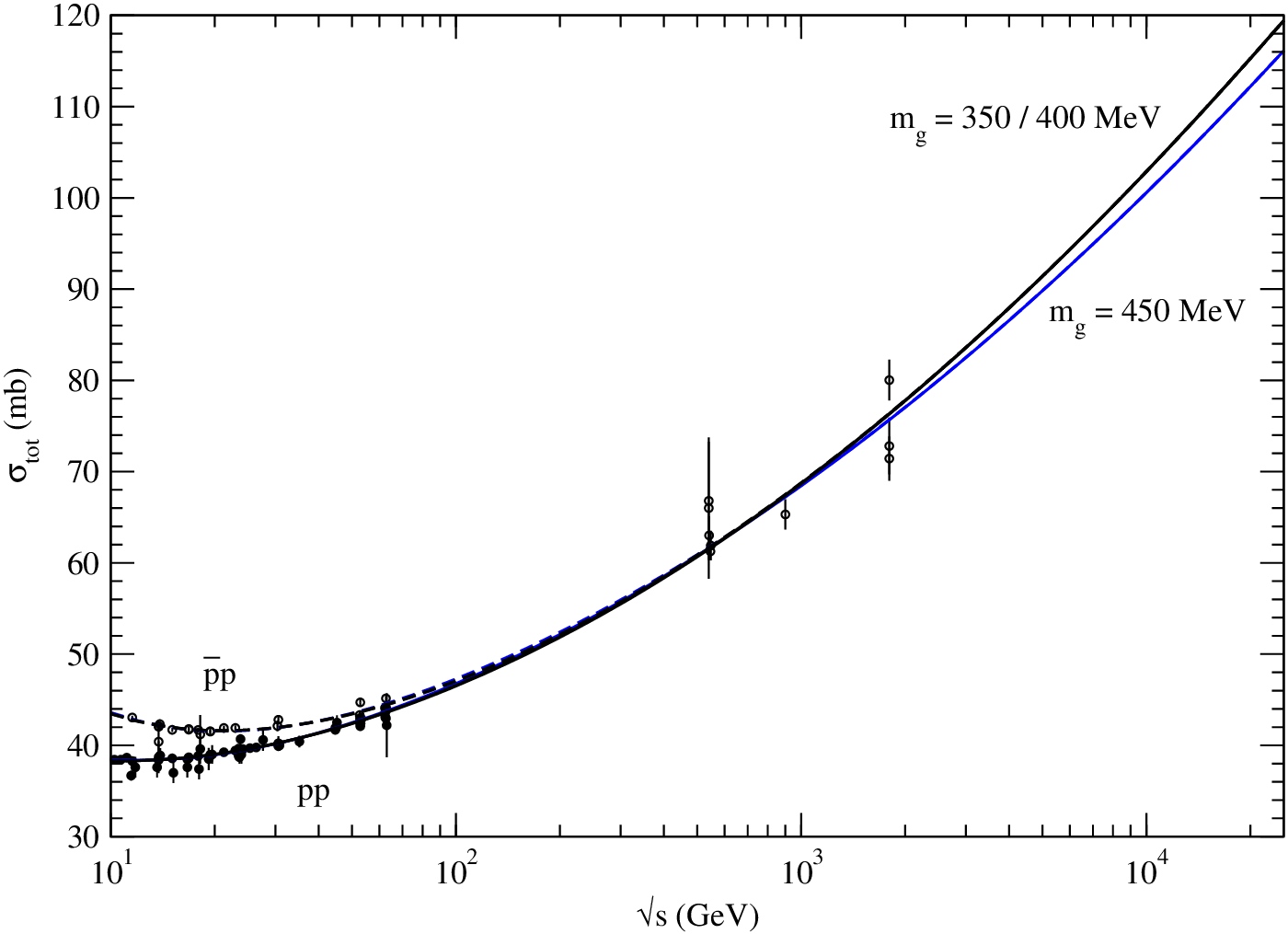}\hspace*{0.2cm}
\includegraphics*[width=8cm,height=7cm]{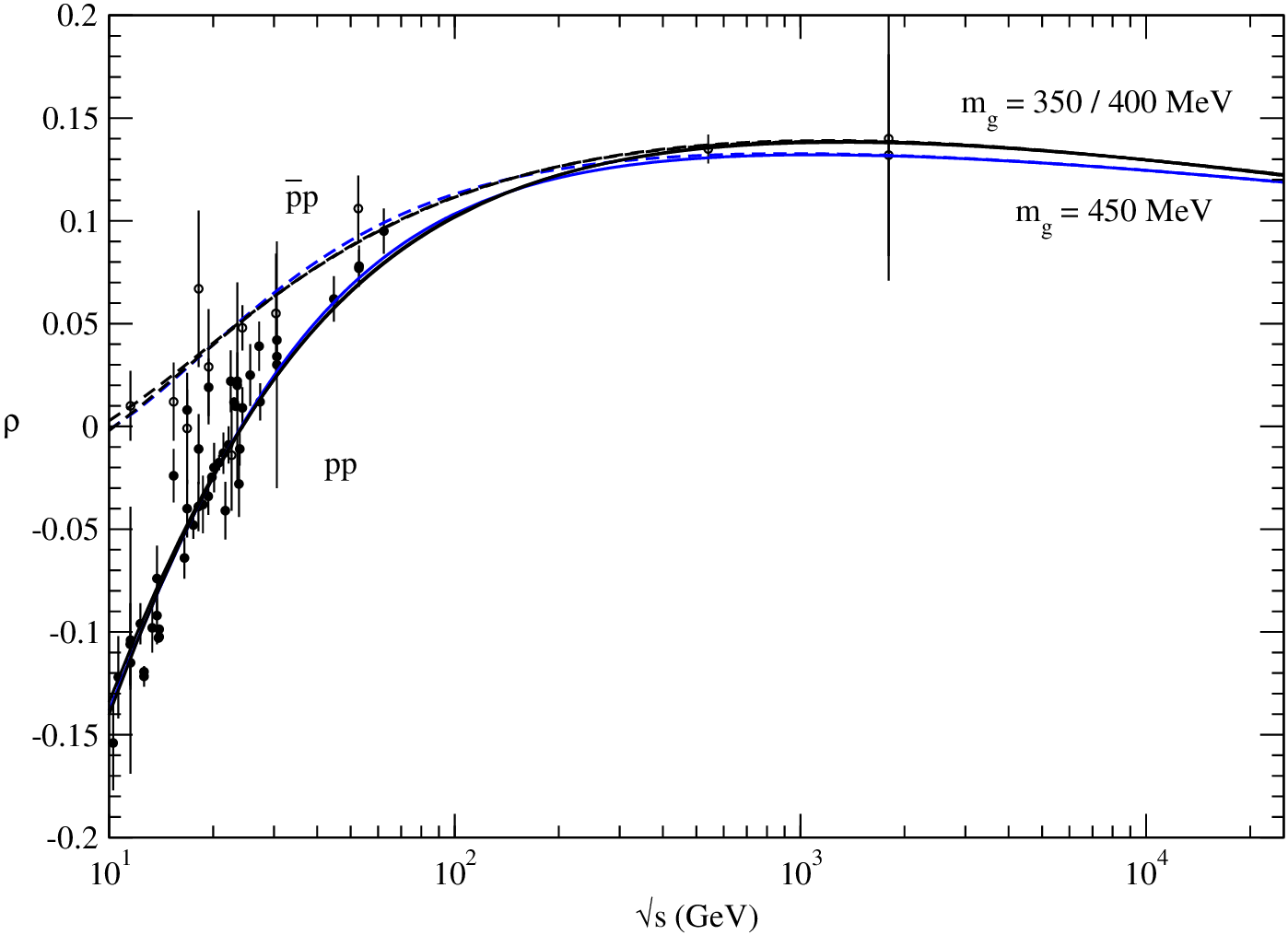}\vspace{0.5cm}
\includegraphics*[width=8cm,height=7cm]{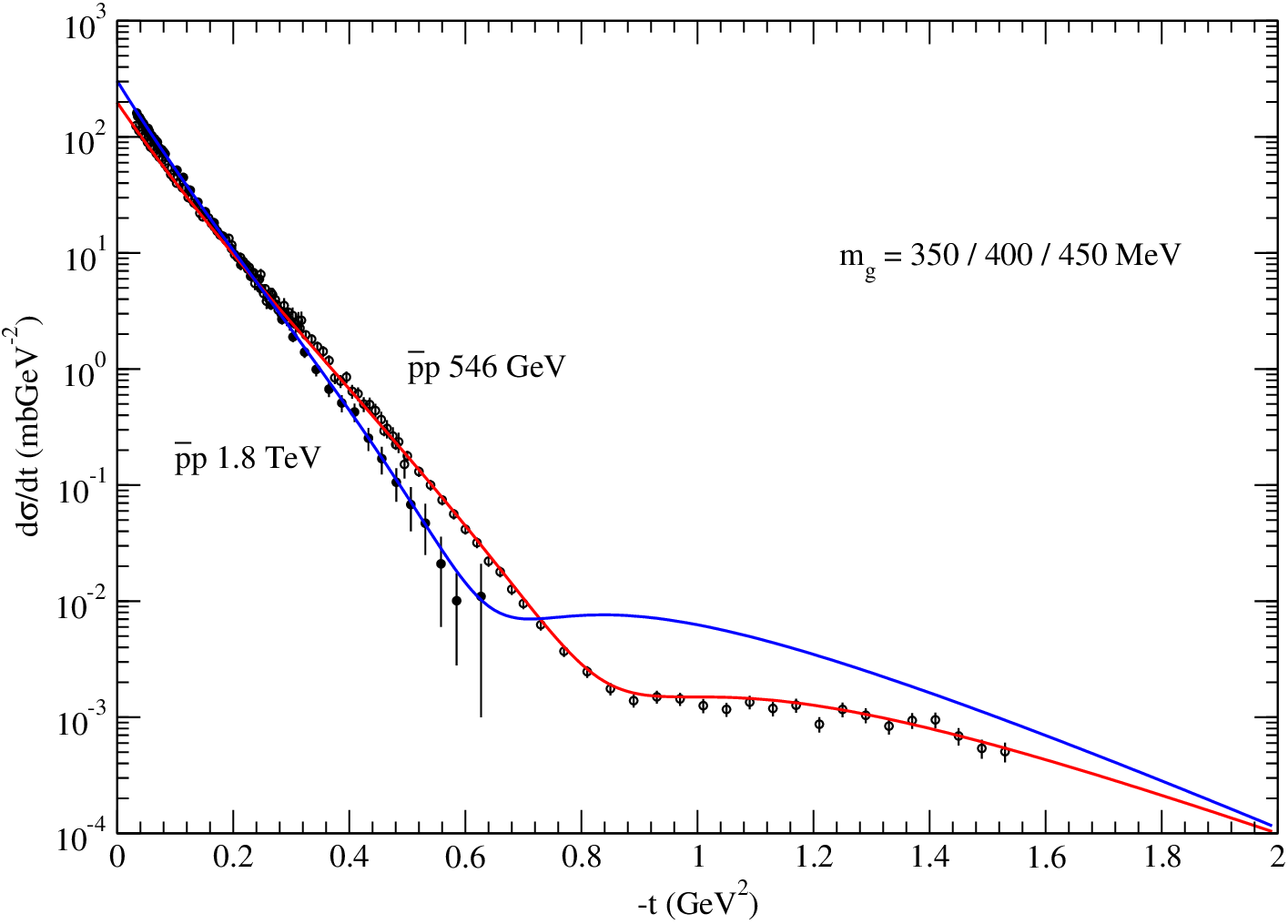}\hspace*{0.2cm}
\includegraphics*[width=8cm,height=7cm]{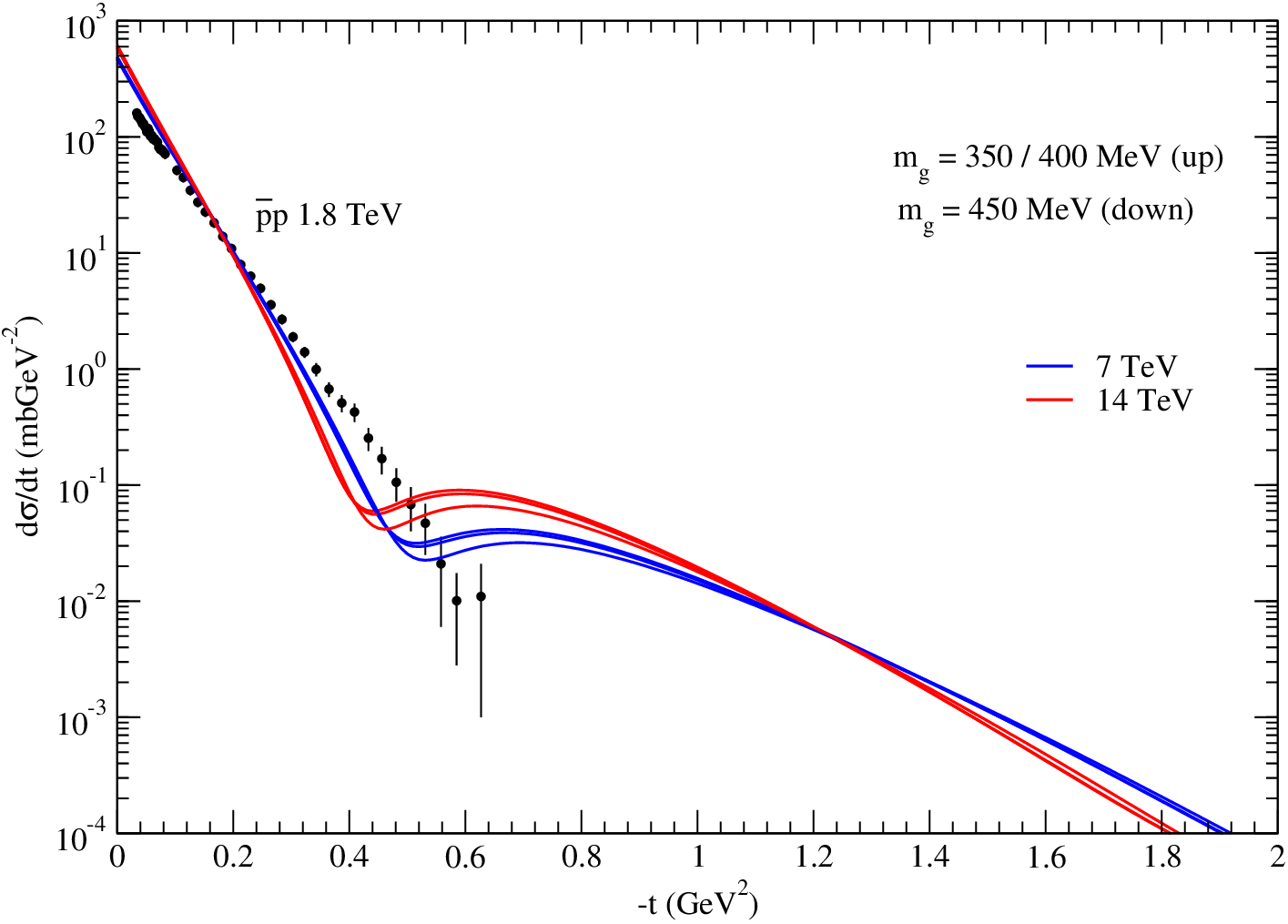}
\end{center}
{\small Fig. 2: Simultaneous fits to $\sigma_{tot}$, $\rho$ and $d\sigma_{el}/dt$ (546 GeV and 1.8 TeV) with $\epsilon$ = 0.08 and $m_{g}$ in the range
350 $-$ 450 MeV; also shown the predictions for LHC at 7 and 14 TeV.}\\\\

\begin{center}
\small{Table 3. Predictions for the $pp$ total cross section and the $\rho$ parameter at the LHC with m$_g$ = 400 MeV
and extreme values of $\epsilon$.}\\\vspace{0.3cm}
\begin{tabular}{ccc}
\hline\hline
$\epsilon$: &0.08&0.09\\
\hline
$\sigma_{tot}(7\ TeV)$ (mb)&96.9 &98.0\\
$\sigma_{tot}(14\ TeV)$ (mb)&108.8&110.6\\
$\rho(7\ TeV)$&0.1321&0.1376\\
$\rho(14\ TeV)$&0.1272&0.1330\\
\hline
\end{tabular} 
\end{center}
\newpage

\begin{center}
\small{Table 4. Predictions for the $pp$ total cross section and the $\rho$ parameter at the LHC with $\epsilon$ = 0.08 
and extreme values of m$_g$.}\\\vspace{0.3cm}
\begin{tabular}{ccc}
\hline\hline
m$_g$(MeV): &350&450\\
\hline
$\sigma_{tot}(7\ TeV)$ (mb)&96.9 &94.9\\
$\sigma_{tot}(14\ TeV)$ (mb)&108.8&106.1\\
$\rho(7\ TeV)$&0.1322&0.1265\\
$\rho(14\ TeV)$&0.1270&0.1225\\
\hline
\end{tabular} 
\end{center}

\section{Conclusions and Final Remarks}
From Table 1 and Figures 1 and 2 the experimental data analysed are quite well described in all cases investigated. For the 
intervals considered  on $\epsilon$ and $m_{g}$ the results are quite similar in the region with available data, 
but that is not the case at higher energies. Tables 3 and 4 show that at the LHC the results present a 3\% of variation 
in the predicted quantities, for the narrow intervals considered in $\epsilon$ and $m_{g}$. That corresponds, for example, to two times the uncertainty 
in predicted values by Block \cite{block01}, where $\epsilon$ = 0.05 and m$_{0}$ = 600 MeV are the ad hoc values used.\\
We conclude that  both parameters,the Pomeron intercept and the mass scale must be carefully investigated and physically justified, in
order to obtain reliable predictions at the LHC energies. We are presently investigating the subject.\\

\textit{Note added.} $\-$ During this workshop we have noticed the recent result by the TOTEM
Collaboration for the pp differential cross section at 7 TeV,
presented by V. Avati and discussed in \cite{totem_cern}. In particular we note that for $\epsilon$ =
0.080 and $m_g$ = 450 MeV (table 2, fig. 2) our model predicts
the dip position at 0.53 GeV$^2$, which is in complete agreement with the
TOTEM result. However, as commented above, we are still investigating
the optimal solution for our free parameters at lower energies and
the final results shall be presented in a forthcoming communication.

\section*{Acknowledgments}
D.A.F. and M.J.M. are thankful to FAPESP (contract Nos. 11/00505-0, 09/50180-0) and A.A.N. to CNPQ for financial support. 
D.A.F. and E.G.S.L. are grateful to Valentina Avati for discussions during the workshop.


\begin{thebibliography}{99}


\bibitem{luna01} E.G.S. Luna, A.F. Martini, M.J. Menon, A. Mihara, and A.A. Natale, Phys. Rev. D {\bf 72}, 034019 (2005).

\bibitem{block01} M. M. Block, Phys. Rep. 436 (2006) 71-215.

\bibitem{block02} M. M. Block, E. M. Gregores, F. Halzen, G. Pancheri, Phys. Rev. D {\bf 60}, 054024 (1999).

\bibitem{cornwall} J. M. Cornwall, Phys. Rev. D {\bf 26}, 1453 (1982); J. M. Cornwall, J. Papavassiliou, Phys. Rev. D 
{\bf 40}, 3474 (1989); J. Papavassiliou and J. M. Cornwall, Phys. Rev. D {\bf 44}, 1285 (1991) 

\bibitem{luna02} E.G.S. Luna, A.A. Natale, and A.L. dos Santos, Phys. Lett. B {\bf 698}, 52 (2011).

\bibitem{lmm} E. G. S. Luna, M. J. Menon, Phys. Lett. B {\bf 565}, 123 (2003); E. G. S. Luna, M. J. Menon, J. Montanha, Nucl.
Phys. A {\bf 745}, 104 (2004).

\bibitem{root} http://root.cern.ch/drupal/; http://root.cern.ch/root/html/TMinuit.html.

\bibitem{totem_cern} G. Antchev \textit{et al}. (TOTEM Collaboration), Europhys. Lett. \textbf{95},  41001 (2011).

\end{thebibliography}
\end{document}